\documentclass[prd,showpacs,preprintnumbers,amsmath,amssymb,superscriptaddress]{revtex4}

\usepackage{color,epsfig}
\usepackage{graphicx}
\definecolor{violet}{rgb}{0.4,0,0.4}
\definecolor{vert}{rgb}{0,0.5,0.0}
\definecolor{navy}{rgb}{0.0,0.0,0.6}
\definecolor{orange}{rgb}{0.8,0.2,0.0}
\definecolor{bleu}{rgb}{0.3,0.0,0.8}

\font\srm=cmr8

\def\be{\begin{equation}}
\def\fe{\end{equation}}

\def\spose#1{\hbox to 0pt{#1\hss}}\def\lta{\mathrel{\spose{\lower 3pt\hbox
{$\mathchar"218$}}\raise 2.0pt\hbox{$\mathchar"13C$}}}  \def\gta{\mathrel
{\spose{\lower 3pt\hbox{$\mathchar"218$}}\raise 2.0pt\hbox{$\mathchar"13E$}}}

\def\Euro{\spose {\lower 2.5pt\hbox{${^{\bf =}}$}}{ C}}

\begin{document}

\title{\bf Elastic properties of anisotropic domain wall lattices}
\author{Richard A. Battye}
\affiliation{Jodrell Bank Observatory, School of Physics and
Astronomy, University of Manchester, Macclesfield, Cheshire SK11
9DL, UK.}
\author{Elie Chachoua}
\affiliation{LuTh, Observatoire de Paris, Meudon 92195, France.}
\author{Adam Moss}
\affiliation{Jodrell Bank Observatory, School of Physics and
Astronomy, University of Manchester, Macclesfield, Cheshire SK11
9DL, UK.}

\date{10/11/2005}
\preprint{}
 
\begin{abstract}
Interest in the elastic properties of regular lattices constructed from domain walls has  recently been motivated by cosmological applications as solid dark energy. This work investigates the particularly simple examples of triangular, hexagonal and square lattices in two dimensions and a variety of more complicated lattices in three dimensions which have cubic symmetry. The relevant rigidity coefficients are computed taking into account non-affine perturbations where necessary, and these are used to evaluate the propagation velocity for any macroscopic scale perturbation mode. Using this information we assess the stability of the various configurations. It is found that triangular lattices are isotropic and stable, whereas hexagonal lattices are unstable. It is argued that the simple orthonormal cases of a square in two dimensions and the cube in three are stable, except to perturbations of infinite extent.  We also find that the more complicated case of a rhombic dodecahedral lattice is stable, except to the existence of transverse modes in certain directions, whereas a lattice formed from truncated octahedra is unstable.
\end{abstract}

\maketitle

\section{Introduction}

The possibility of an effective negative macroscopic pressure due to a medium constituted from a lattice of cosmic strings, and {\it a fortiori} domain walls, has motivated interest in such structures as the mechanism~\cite{BS98,BBS99} which might account for the observed acceleration of the universe~\cite{BM05}. A crucial issue that arises is whether such structures can be sufficiently rigid to offset the negative pressure, and behave as a stable elastic solid. A recent investigation~\cite{BCCM05} has shown that a randomly orientated, and thus macroscopically isotropic, distribution of cosmic strings or membranes will indeed be sufficiently rigid if only affine deformations are considered. It was explained that this should apply in the case where the junctions are of even-type (X-type or $\star$-type), that is, when an even number of elements (strings or walls) meet at each junction.

The purpose of the present work is to initiate a study of the macroscopic elastic properties, and hence the local stability properties, of more realistic models in which the solid is formed from a regular lattice structure, which is naturally anisotropic. Since wall lattices, with $P/\rho=-2/3$ in 3 dimensions, are favoured by observations we will concentrate on this possibility, but our methods could equally well be applied to the string case with $P/\rho=-1/3$. We will consider the cases of wall lattices with triangular, hexagonal and square symmetry in two dimensions, and three dimensional lattices with cubic symmetry which can be created from the Wigner-Seitz cells of simple cubic, face-centred cubic and body-centred cubic lattices. The unit cells of these primitive lattices are the simple cube, the rhombic dodecahedron and the truncated octahedron respectively.

The important issue which we need to address is that some of the regular lattices we would like to consider contain junctions that are of odd-type (an odd number of elements meeting at a junction). Y-type junctions of this kind appear in, for example, models with hexagonal symmetry where three walls meet at 120 degrees, and are ubiquitous in three dimensional models. As was pointed out in ref.~\cite{BCCM05} such models require consideration non-affine deformations to maintain the equilibrium condition at the junction, violating the assumptions of the earlier calculation. Typically such non-affine deformations will lead to a reduction in the rigidity when compared to the affine case, making the structures less stable. In addition to this, it will also be necessary to introduce an extra shear modulus to take into account the anisotropic nature of lattices with square and cubic symmetries.

First, we will establish the general conditions for stability in the isotropic case and cubic cases in general dimensions. This formalism can then be applied to the cases of triangular and hexagonal symmetry in two dimensions. It is shown that such systems are effectively isotropic, but that the triangular case is stable, whereas the hexagonal case is not. We then apply this to simple orthonormal cases of a square in two dimensions and simple cubic system in three dimensions. We find that such systems are anisotropic and have longitudinal zero modes corresponding to a finite number of specific directions. We argue that such structures are only unstable to perturbations of infinite extent. Finally, we consider wall lattices whose unit cells are the truncated octahedron and the rhombic dodecahedron. The former is unstable, but the latter is only unstable to transverse zero modes in certain directions.

\section{Formalism}

\subsection{Quasi Hookean ansatz}

The formulation of a perfectly elastic solid model is based on a material base space which is the quotient of the spacetime by the congruence of the material worldlines. The $D$-dimensional, time dependent metric induced on this base space is denoted $\gamma_{ab}$. For the purposes of our discussion it will be sufficient to use models which are characterised by a {\it quasi Hookean} equation of state~\cite{CQ72}. This implies that the energy density can be expressed in terms of the  {\it rigidity tensor}, $\check\Sigma^{abcd}$, as 
\be \rho=\check\rho+\frac{_1}{^2}\check{\mit\Sigma}^{abcd}s_{ab}s_{cd}
\,.\label{01}\fe
The overhead check symbol is used to indicate quantities which depend only on the value of the conserved number density, $n$. The quantity $\check\rho$ is the minimum value taken by the mass-energy density, $\rho$, for the equivalent value $n$. It is important to distinguish between $\gamma_{ab}$ and its value $\check \gamma_{ab}$ in the reference state for which this minimum is attained.
A quadratic form, such as (\ref{01}), can be expected to be valid 
as a realistic approximation for sufficiently small values of the {\it 
constant volume shear tensor} $s_{ab}$ which is defined by
\be s_{ab}={1\over 2} \left(\gamma_{ab}-\check\gamma_{ab}\right)
\, .\label{02}\fe

The functional dependence of $\check\rho$ on $n$ determines corresponding reference state value of the pressure tensor, $\check P^{ab}$, and also of the bulk modulus, $\check\beta$. These are given by 
\be \check P^{ab}=\check P\check\gamma^{-1\, ab}\qquad
\check P=n\frac{{\rm d}\check\rho}{{\rm d} n}-\check\rho\, ,
\qquad \check\beta=n \frac{{\rm d}\check P}{{\rm d} n}
\, ,\label{00}\fe
in terms of the inverse metric $\check\gamma^{-1\, ab}$ induced on the base space. The rigidity tensor on the material base space has the following symmetry 
conditions
\be \check{\mit\Sigma}^{abcd}=\check{\mit\Sigma}^{(ab)(cd)}=
\check{\mit\Sigma}^{cdab}
\, .\label{03}\fe

The fact that $\gamma_{ab}$ and $\check\gamma_{ab}$ must have the 
same determinant (due to the conservation of volume) 
implies that on the $D$-dimensional material base the 
symmetric tensor $s_{ab}$ will have only $(D^2+D-2)/2$ (instead of $D(D+1)/2$) independent components, and more specifically that to first order it will be trace free with respect to either the actual metric $\gamma_{ab}$ or the 
 reference metric $\check\gamma_{ab}$. Therefore, it is necessary to impose 
a restriction to completely fix the specification of 
$\check{\mit\Sigma}^{abcd}$. This can be most conveniently achieved~\cite{CQ72} by requiring that it be trace free with respect to the reference metric
{\be \check{\mit\Sigma}^{abcd}\check\gamma_{cd}=0\, .\label{04}\fe}

\subsection{Characteristic propagation equation}

We shall be considering systems characterised by the absence of {\it compressional distortion} which means that the only effect of a change in $n$ on the reference metric $\check\gamma_{ab}$ is to multiply it by a simple conformal factor. Under these circumstances the complete elasticity tensor $\check E^{abcd}$ in the reference state will differ from the rigidity tensor $\check{\mit\Sigma}^{abcd}$ by an isotropic  contribution determined just by the dependence of the reference density $\check\rho$ on $n$. This is given by 
\be \check E^{abcd}=\check{\mit\Sigma}^{abcd}+
2\check P\check\gamma^{-1\, c(a} \check\gamma^{-1\, b)d}
+(\check\beta-\check P)\check\gamma^{-1\, ab}\check\gamma^{-1\, cd}
\,.\label{06}\fe

This leads to an expression of the form
\be \check A^{abcd}=\check{\mit\Sigma}^{abcd}+\check\beta\check
\gamma^{-1\, ab}\check\gamma^{-1\, cd}-\check P
\check\gamma^{-1\, a[b} \check\gamma^{-1\, c]d}\,, \label{17}\fe
for the relativistic Hadamard elasticity tensor~\cite{C83} that is needed 
for the formulation of the characteristic equation governing the 
propagation of small perturbations. It follows~\cite{C72,CCC05} that the characteristic equation for the speed $\upsilon$ of a mode with polarisation direction $\iota_a$ propagating in the direction specified by a unit space vector $\nu^a$ will be given by
\be\bigg[\upsilon^2(\check\rho+\check P)\check
\gamma^{-1\, ab}-Q^{ab}\bigg]\iota_a=0\,,\label{18}\fe
where
\be Q^{ab}=\check{\mit\Sigma}^{acbd}\nu_c\nu_d +\check\beta\nu^a\nu^b
\, .\label{20}\fe
The eigenmodes of this equation will yield the normal modes of the system and will allow us to establish stability conditions.

\subsection{Isotropic case}

A system of the kind we have been considering will evidently reduce to 
one of simple perfect fluid type if and only if the rigidity tensor
$\check\Sigma^{abcd}$ is zero, a condition that can be seen to be 
equivalent to that of vanishing of the scalar rigidity modulus $\check\mu$
as defined~\cite{CQ72} by the formula
{\be \check\mu^2=\frac{1}{2({\srm D}+2)({\srm D}-1)}
\check {\mit\Sigma}^{abcd}\check {\mit\Sigma}^{efgh}\check\gamma_{ae}
\check\gamma_{bf}\check\gamma_{cg}\check\gamma_{dh}
\,.\label{09}\fe}

The normalisation in (\ref{09}) is determined by the requirement of 
consistency with traditional terminology~\cite{LL} in the ordinary 
isotropic case, for which usual convention is that the rigidity term 
in (\ref{01}) should simply be given by
{\be \frac{_1}{^2}\check{\mit\Sigma}^{abcd}s_{ab}s_{cd}=
\check\mu\, \check\gamma^{-1\, c(a} \check\gamma^{-1\, b)d}
s_{ab}s_{cd}=\check\mu s_{ab}s^{ab}\, ,\label{05}\fe}
so that the corresponding expression for the isotropic rigidity tensor 
will be 
{\be \check{\mit\Sigma}^{abcd}= 2\check\mu\left(
\check\gamma^{-1\, c(a} \check\gamma^{-1\, b)d}-\frac{1}{\srm D}
\check\gamma^{-1\, ab}\check\gamma^{-1\, cd}\right)
\label{08}\fe}
This means that the non vanishing Cartesian components of 
the rigidity tensor $\check{\mit\Sigma}^{abcd}$ in (\ref{01}) will just 
be of three kinds. There will be the mixed kind specified by a pair of 
axes,  ${\mit\Sigma}^{_{1\,2\,1\,2}}$, 
the crossed kind also specified by a pair of axes, ${\mit\Sigma}^{_{1\,1\,2\,2}}$, and finally the pure kind specified by a single axis, ${\mit\Sigma}^{_{1\,1\,1\,1}}$. According to (\ref{08}) the values of
these components will be given by
{\be {\mit\Sigma}^{_{1\,2\,1\,2}}=\check\mu\, ,\hskip 1 cm
{\mit\Sigma}^{_{1\,1\,2\,2}}=-\frac{2\check\mu}{\srm D}
\, , \hskip 1 cm {\mit\Sigma}^{_{1\,1\,1\,1}}=
\frac{2({\srm D}-1)\check\mu}{\srm D}\, .\label{14}\fe}

It can be seen from (\ref{08})  that, in a medium of this isotropic 
kind, the polarization eigenvector $\iota_a$ in (\ref{18}) will have one 
of two distinct types. It may be orthogonal to the propagation direction 
$\nu^a$, $\iota_a\nu^a=0$, in which case the corresponding transverse
velocity eigenvalue $\upsilon_{\rm T}$ will be given simply by
\be \upsilon_{\rm T}^{\,2}={\check \mu\over \check\rho+\check P}
\, .\label{21}\fe
Alternatively the polarisation direction  $\iota_a$ may be parallel to 
the propagation direction $\nu^a$, in which case the corresponding 
longitudinal velocity eigenvalue $\upsilon_{\rm L}$ will be given by
\be \upsilon_{\rm L}^{\,2}={
D\check\beta+2(D-1)\check\mu\over \check D(\rho+\check P)}
\,.\label{23}\fe
There are $D-1$ orthogonal directions in the eigenspace with velocity $\upsilon_{\rm T}$ and one with $\upsilon_{\rm L}$. In the case where the equation of state is given by $\check P=w\check \rho$ the two sound speeds are given by
\be \upsilon_{\rm T}^{\,2}={\check\mu\over (1+w)\check\rho}\,,\qquad \upsilon_{\rm L}^{\,2}=w+{2(D-1)\check\mu\over D(1+w)\check\rho}\,.\fe

We shall be concerned with cases where the medium will 
consist of a relativistic system of Dirac-Goto-Nambu type branes. In such a system the reference state can be described by a polytropic equation of state with index $\gamma$, 
so that
\be \check P=w\check\rho\, ,\qquad \check \beta=\gamma w\check 
\rho\, ,\qquad w=\gamma-1\, ,\label{24}\fe
where $\gamma$ is the ratio of the codimension to the space dimension,
so that for the generic p-brane case one obtains
\be \gamma=\frac{({\srm D}-p)}{\srm D}\, ,\qquad
w= -\frac{p}{\srm D}\, .\label{26}\fe
This means that both for strings, as characterised by $p=1$, and walls, 
as characterised by $p={\srm D}-1$ we shall always have
\be \check\beta=-\frac{({\srm D}-1)\check\rho}{{\srm D}^2}
\, ,\label{25}\fe
for arbitrary space dimension ${\srm D}$. Thus, in the ordinary 3 
dimensional case  the bulk modulus will be given by $\check\beta
=-2\check\rho/9$, not only for strings, with $\gamma=2/3$ and  
$w=-1/3$, but also for walls, with $\gamma=1/3$ and $w=-2/3$. In the  
2 dimensional case, for which there is no need to distinguish between 
strings and walls, one will simply have $\check\beta=-\check\rho/4$ 
with $\gamma=1/2$ and  $w=-1/2$.  

In all these cases, it can be seen that for a system of the isotropic
kind characterised by (\ref{08}) the corresponding propagation speeds of
transverse and longitudinally polarised perturbation modes will be given 
by
\be \upsilon_{\rm T}^{\,2} =\frac{\check\mu}{\gamma\check\rho}
\, ,\hskip 1 cm \upsilon_{\rm L}^{\,2}=\frac{2({\srm D}-1)}
{\gamma{\srm D}}\left(\frac{\check\mu}{\check\rho}-\frac{1}{2\srm D}
\right)\, .\label{27}\fe

\subsection{Stability and causality in the isotropic case}

In order for the system to be stable and respect causality, we require that the two sound speeds satisfy $0\le\upsilon_{\rm T}^{\,2}\le 1$ and $0\le\upsilon_{\rm L}^{\,2}\le 1$, since otherwise there will be modes propagating at speeds faster than the speed of light, or there will be modes whose amplitude grows exponentially on a timescale proportional to its wavelength. In the general case of $\check P=w\check \rho$ this implies that 
\be 
0\le {\check\mu\over\check\rho}\le 1+w\,,\qquad -{Dw(1+w)\over 2(D-1)}\le {\check\mu\over\check\rho}\le {D(1-w^2)\over 2(D-1)}\,,
\fe
where we have assumed that $1+w>0$.

For systems described by (\ref{24},\ref{26},\ref{25},\ref{27}), the stability criterion can be seen to be that the rigidity modulus is not less
 than a critical minimum value given
\be \check\mu=\frac{\check\rho}{2{\srm D}}\, ,\label{19}\fe
which implies that the critical minimum value is $\check\mu=\check\rho/6$
for  ${\srm D}=3$ and $\check\mu=\check\rho/4$ for  ${\srm D}=2$.
For an isotropically randomised distribution of strings or membranes considered in our previous work~\cite{BCCM05} this will be satisfied since it was shown that the relevant  rigidity modulus will be given for ${\srm D}=3$ by $\check\mu= 
4\check\rho/15$ , and it can easily be verified that an an analogous 
calculation for the simpler case  ${\srm D}=2$ gives $\check\mu= 3\check\rho/8$. Moreover, the condition for causal wave propagation is that
\be
\frac{\check\mu}{\check\rho} \leq \gamma\,,\qquad {\check\mu\over\check\rho}\leq {1\over 2D}+{\gamma D\over 2(D-1)}\,.
\fe
In cases of present interest with $D=2$ and $D=3$, it can be seen that the first of these two formula yields the strongest constraint, but this need not be the case for larger values of $D$. It is clear that these are also satisfied by the model discussed in ref.~\cite{BCCM05}

We should note that we have allowed for the possibility of zero modes in our analysis above, that is, we have classed the cases of $\upsilon_{\rm T}=0$ and $\upsilon_{\rm L}=0$ as stable. This is something which is somewhat delicate and more consideration of that particular context is necessary. A familiar example of such a mode is provided by the ordinary barotropic perfect fluid, as characterized by a strictly positive bulk modulus, $\check\beta$, and vanishing rigidity modulus, $\check\mu$. In this case $\upsilon_{\rm L}^{\,2}>0$, but $\upsilon_{\rm T}=0$. This indicates that transverse shear modes do not propagate and are not subject to exponential growth. In a non-expanding background, they are subject to unbounded linear growth and therefore can be described as marginally unstable. However, in an expanding background they are expected to be subjected to either power law growth or decay dependent on the details of the situation.

Furthermore, it is worthwhile to reflect on how the existence, in
principle, of such marginally unstable modes can be compatible with
our everyday experience of the stability of ordinary static fluid
states. The key to the paradox is the realisation that the modes in
question are plane waves of infinite extent.  For such an instability to grow,  there must exist growing perturbations that are
initially confined within a locally bounded support region.  The
existence of marginally unstable  modes of plane wave type and thus
with unbounded support is not always sufficient to provide unstable
wave-packet combinations with bounded support.

An example in which there are no exponentially unstable modes, but in
which (unlike the perfect fluid case) there actually are sufficient
marginally unstable plane wave type  modes to form growing
perturbations with local support is provided by the hexagonal lattice
model discussed in the next section. This Y-type example is to be
contrasted with the X-type cases discussed in subsequent  sections,
namely the square and cubic lattice models, for which some zero modes
do exist but, as in the fluid case, are not enough to give rise to an
effective locally supported instability.

\subsection{Simple anisotropic cases}

The possible anisotropic forms for the rigidity tensor can be classified for $D=2$ and $D=3$ in terms of the Bravais lattices~\cite{LL} extensively studied in crystallography. We will consider here the simplest of these cases, that of the orthonormal type, meaning 
lattices of simple square type for ${\srm D}=2$ and  lattices of simple 
cubic type for ${\srm D}=3$, not to mention hypercubic types for larger 
values of ${\srm D}$  that might be worth considering in view of the 
current interest in higher dimensional cosmological models. It is well documented in the literature that, due to their symmetry under permutation of the axes, models of this type have non vanishing Cartesian components of the rigidity tensor $\check{\mit\Sigma}^{abcd}$ in (\ref{01}) of three kinds, with just two independent rigidity parameters. We will suggestively designate these parameters as $\check\mu_{\rm T}$ and
$\check\mu_{\rm L}$. There will be the mixed kind specified by a 
pair of axes, for example ${\mit\Sigma}^{_{1\,2\,1\,2}}$, the crossed kind also specified by a pair of axes, for example
$ {\mit\Sigma}^{_{1\,1\,2\,2}}$ and finally the pure kind specified 
by a single axis, for example
${\mit\Sigma}^{_{1\,1\,1\,1}}$. The requirement  (\ref{04}) that the 
rigidity tensor be trace free implies that the value of a component 
of the pure type must be given in terms of that of a component of the 
crossed type by a relation of the form
{\be  {\mit\Sigma}^{_{1\,1\,1\,1}}=(1-{\srm D}){\mit\Sigma}^{_{1\,1\,2\,2}}
\, .\label{45}\fe}
In order to ensure that they both go over to the ordinary scalar rigidity
modulus $\check\mu$ in the isotropic limit as given by (\ref{14}) the 
choice of the independent rigidity coefficients
$\check\mu_{\rm T}$ and $\check\mu_{\rm L}$ can thus be
conveniently fixed by the specifications
\be {\mit\Sigma}^{_{1\,2\,1\,2}}=\check\mu_{\rm T}\, ,\hskip 1 cm
{\mit\Sigma}^{_{1\,1\,2\,2}}=-\frac{2\check\mu_{\rm L}}{\srm D}
\, , \hskip 1 cm {\mit\Sigma}^{_{1\,1\,1\,1}}=
\frac{2({\srm D}-1)\check\mu_{\rm L}}{\srm D}\, ,\label{15}\fe
which entail that the rigidity scalar $\check\mu$ defined by (\ref{09}) 
will be given, as a weighted root square mean, by
\be \check\mu^2= \frac{{\srm D} \check\mu_{\rm T}^{\,2}
+2\check\mu_{\rm L}^{\,2}}{{\srm D} +2}\,. \label{16}\fe
Moreover, we define $\Delta\check\mu=\check\mu_{\rm T}-\check\mu_{\rm L}$ which quantifies the anisotropy of the lattice.

Writing $\nu^a={\rm cos}\theta_a$, where $\theta_a$ is the angle
between the propagation direction and the $x^i$ axis, it can
be seen from the formula (\ref{20}) that the diagonal components 
of the matrix $Q^{ab}$ in the characteristic equation (\ref{18})
can be written as (no summation)
\begin{equation}
Q^{ii}=\left( \frac{2(D-1)}{D} \check{\mu}_{L} + \check{\beta} \right) \cos^{2} \theta_{i} + \check{\mu}_{T} \sin^{2} \theta_{i},
\label{qii}
\end{equation}
for $1\le i\le D$ and the off-diagonal components are given by
\begin{equation}
Q^{ij} =\left( -\frac{2 \check{\mu}_{L}}{D} + \check{\mu}_{T} + \check{\beta}\right) \cos \theta_{i} \cos \theta_{j}\,,
\label{qij}
\end{equation}
for $i\ne j$.

\subsection{Stability and causality in the general orthonormal case}
\label{anisostable}
In an anisotropic system, such as the cubic type considered here, the propagation speed  of perturbations modes are typically spatially dependent making the discussion of stability more complicated than in the isotropic case. The characteristic propagation equation can be solved by solving an eigenvalue problem for $\lambda=(\check\rho+\check P)\upsilon^2$. In general this is a complicated polynomial equation of degree $D$, with the coefficients depending on $\cos\theta_{\rm i}$ for $i=1,..,D$. The characteristic equation for general $D$ is complicated; here  we present it for the cases of specific interest $D=2$,
\be 
3 \lambda^2 -  (4 \check\mu_{\rm L} + 3 \beta + 3 \check\mu_{\rm T}) \lambda  + \check\mu_{\rm T} ( 3 \beta + 4 \check\mu_{\rm L}) + 4 B_{2} (\check\mu_{\rm L}-\check\mu_{\rm T})(\check\mu_{\rm L}+ 3 \beta) =0\,,
\label{ch2}
\fe
and $D=3$, 
\be
\begin{array}{c}
3 \lambda^{3} - ( 4 \check\mu_{\rm L} + 3 \beta + 6 \check\mu_{\rm T} )  \lambda^{2}  \\ + \left \{ \check\mu_{\rm T} (8 \check\mu_{\rm L}  + 6 \beta + 3 \check\mu_{\rm T} ) + 4 ( \check\mu_{\rm L} - \check\mu_{\rm T} )( \check\mu_{\rm L}  + 3 \beta) B_{3} \right \} \lambda  \\
 - \check\mu_{\rm T}^{2} \left( 4 \check\mu_{\rm L} + 3 \beta \right) - 4 \check\mu_{\rm T} ( \check\mu_{\rm L} - \check\mu_{\rm T} )(\check\mu_{\rm L}  + 3 \beta ) B_{3} - 4 (\check\mu_{\rm L}-\check\mu_{\rm T})^{2}( \check\mu_{\rm T} + 3 \beta ) C  = 0 \,,
\label{ch3}
\end{array}
\fe
where $B_{2}=\cos^{2} \theta_{1} \cos^{2} \theta_{2}$, $B_{3}=\cos^{2} \theta_{1} \cos^{2} \theta_{2}  + \cos^{2} \theta_{1} \cos^{2} \theta_{3} + \cos^{2} \theta_{1} \cos^{2} \theta_{3}$ and $C=\cos^{2} \theta_{1} \cos^{2} \theta_{2} \cos^{2} \theta_{3}$

For specific value of the rigidity coefficients $\check\mu_{\rm L}$ and $\check\mu_{\rm T}$ the $D$ solutions of the characteristic equation can be computed and unlike the isotropic case these will, in general, be distinct and be dependent on direction. 

We have been able to solve (\ref{ch2}) in general, 
\be 
\upsilon^{2}_{_{(1,2)}} = { (4 \check\mu_{\rm L} + 3 \beta + 3 \check\mu_{\rm T}) \pm \sqrt{(4 \check\mu_{\rm L} + 3 \beta - 3 \check\mu_{\rm T} )^{2} -  48 B_{2} (\check\mu_{\rm L}-\check\mu_{\rm T})  (\check\mu_{\rm L} + 3 \beta)} \over 6(\check\rho+\check P)}\,.
\fe 
which yields stability conditions $\check\mu_{\rm L}\ge\check\rho/4$ and $\check\mu_{\rm T}\ge\check\rho/4$ for $\check\beta=-\check\rho/4$ as suggested by (\ref{25}).
However, in the case $D=3$ and, indeed, for larger values of $D$ we have been unable to solve the characteristic equation analytically. Of interest when considering the stability are the regions of the $(\check\mu_{\rm L},\check\mu_{\rm T})$ space where $\lambda=0$ is a solution. This will require the coefficient of $\lambda^0$ in the characteristic equation to be zero. For general $D$, this coefficient is minimized for two specific directions: (1) $D$ axial directions corresponding to $\theta_{i}=0$ for each $1\le i\le D$ and $\theta_{j}= \pi/2$ for $i \neq j$,  and (2) a single non-axial direction $\cos{\theta_{i}}=1/D$ for all $i=1..D$.

The characteristic equation for general $D$ can then be factorized. In the axial directions the characteristic equation is given by
\be
\bigg(\check{\mu}_{\rm T}-\lambda\bigg)^{D-1}\bigg(2(D-1) \check{\mu}_{\rm L} + D \check{\beta} - D \lambda\bigg)=0,
\fe
and in the non-axial direction it is given by
\be \label{nonaxial}
\bigg(2 \check{\mu}_{\rm L}+(D-2) \check{\mu}_{\rm T}-2\lambda\bigg)^{D-1}\bigg(2(D-1) \check{\mu}_{\rm T} + D \check{\beta} - 2 \lambda\bigg)=0.
\fe
If we assume that $\check\beta$ is given by (\ref{25}), then one can ensure that all propagation modes are stable in these directions, if the rigidity moduli are not be lower than critical values given by 
\be
\check{\mu}_{\rm L}=\check{\mu}_{\rm T}=\frac{\check{\rho}}{2D},
\label{stmod}
\fe
which is reminiscent of the isotropic case and is exactly the same as for $D=2$.

Therefore, we have shown that if the rigidity moduli are given by (\ref{stmod}) then there are non-propagating modes. We have not shown rigorously that these are the unique points at which there is a zero eigenvalue solution to the characteristic equation, nor have we shown that that there is an negative eigenvalue, and hence an unstable mode below this point. However, we have checked this numerically by computing the lowest eigenvalue for a discrete set of points in the $(\check\mu_{\rm L},\check\mu_{\rm T})$ plane, confirming that the conditions for stability are that
\be{\check\mu_{\rm L}\over\check \rho}\ge {1\over 2D}\,,\qquad {\check\mu_{\rm T}\over \check \rho}\ge {1\over 2D}\,.\fe
Similarly, one can show that causality requires that $\check\mu_{\rm L}/\check\rho\le \gamma$ and $\check\mu_{\rm T}/\check\rho\le \gamma$ for $D=2$ and $D=3$.

\section{Shear Moduli for Triangular and Hexagonal lattices in 2D}

A particularly instructive -- simple, but non-trivial -- example is 
provided by the case of a regular lattice of 
Dirac-Nambu-Goto type domain wall boundaries separating hexagonal 
domains in 2 dimensions. It is illuminating to compare it with the
more artificial and technically more trivial case of its dual lattice, the simple triangular lattice.

\begin{figure}
\centering\epsfig{figure=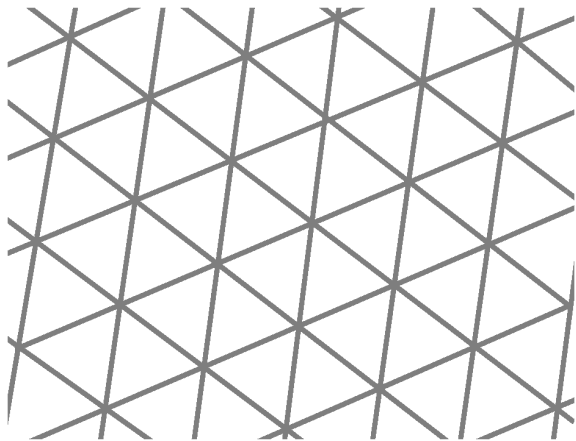, height=6 cm}
\caption{ The case of a triangular cosmic string
lattice with even type ``asterisk'' form junctions in 2 dimensions,
a system that is quite stable, as it admits no localised
 perturbations with negative or zero energy.
}
\label{Fig1}\end{figure}

In both the triangular case, as illustrated in Figure 1, and the 
hexagonal case, as illustrated in Figure 2, the length and direction 
of each boundary section is represented by one, or other, of three 
distinct 2 dimensional vectors, with components $b_{_{(0)}}^{\,a}$,  
$b_{_+}^{\,a}$,  $b_{_-}^{\,a}$ which branch off from each other at $120^{\circ}$ angles. Without loss of generality the space coordinates 
can be chosen in such a way that in the unperturbed state
the first of these is a unit vector negatively oriented along the
$x^{_1}$ axis, so that the complete set will be given by
{\be b_{_{(0)}}^{\,a}\ \leftrightarrow \ \{-1,0\}\, ,\hskip 1 cm
b_{_\pm}^{\,a}\ \leftrightarrow\ \left\{\frac{1}{2},\pm\frac{\sqrt 3}{2}
\right\}\, .\label{30}\fe}

It is evident that such boundary segments might be fitted together to 
form a regular periodic triangular lattice, with any two of the 
three boundary segment vectors as a fundamental dyadic pair of independent 
symmetry generators, and with junctions of the, six fold, even ``asterisk'' type ($\star$-type).  Such a junction between  an even number, in this case 6, of 
terminating segments can also be described as the crossing 
of half that number, in this case 3, uninterrupted boundaries.
It follows that its equilibrium will be preserved by a simple affine
linear space deformation.

Although this network appears to be anisotropic, it is in fact not the case. To see this, let us evaluate which are the non-vanishing components 
of the rigidity tensor. Since the system has an invariance with respect to rotations, the easiest way is to look at the problem in the complex
 coordinates as defined by 
\begin{eqnarray}
\xi= x + i y\,,\qquad\eta= x- i y\,,
\end{eqnarray}
where (x,y) are the Cartesian coordinates. In these new coordinates, a rotation of angle $\alpha$ comes back to the simple transformation
\begin{eqnarray}
\xi \mapsto e^{i \alpha} \xi\,,\qquad \eta \mapsto e^{-i \alpha} \eta
\end{eqnarray}

Since the system is invariant under rotations of angle $\alpha=\frac{\pi}{6}$ the only non vanishing components of the rigidity tensor will be the ones with 
as many  $\xi$ as $\eta$ , that is  $\Sigma^{\xi \xi \eta \eta}$ and  $\Sigma^{\xi \eta  \xi \eta}$. However, the fact that  $\Sigma^{abcd}$ is traceless
 brings back the non vanishing components to one, which is easily seen to be  $\Sigma^{\xi \xi \eta \eta}$. This implies that the system is isotropic.
It is to be remarked that the previous argument applies as well to the Y-type junction hexagonal 2D lattice (which is invariant under rotations of $\alpha=\frac{\pi}{3}$ ). By evaluating the effect of the affine deformation on the triangular lattice in a similar way to our preceding work~\cite{BCCM05}, 
it can be shown that the rigidity modulus is given by the same formula as for an isotropic random distribution discussed earlier, namely 
\be \check\mu=\frac{3\check\rho}{8}\, .\label{28}\fe
The fact that this rigidity value is substantially larger than the
relevant two dimensional critical value, $\check\mu=\check\rho/4$ ensures 
that such an even type triangular lattice will be absolutely stable.
It can be seen from (\ref{27}) that the corresponding propagation
speeds for transverse and longitudinally polarised perturbation
modes are given by
\be \upsilon_{\rm T}^2=\frac{3}{4}
\, ,\hskip 1 cm \upsilon_{\rm L}^{\,2}=\frac{1}
{4} \, .\label{29}\fe
While this stability feature may be considered to be an
advantage from an engineering point of view, on the other hand from 
a more purely physical point of view, such a triangular lattice has 
the drawback of being relatively difficult to obtain in a 
natural context.  This contrasts with the case of a hexagonal 
lattice such as can be formed naturally by a 
system with three symmetrically related vacuum states.

A regular hexagonal lattice can be constructed by fitting together boundary segments of the same three kinds as in the triangular case. However, instead of 
being directly identified with any two individual members of the set
three boundary segment vectors  $b_{_{(0)}}^{\,a}$,  $b_{_+}^{\,a}$,  
$b_{_-}^{\,a}$ as in the triangular case, the fundamental dyad of 
periodicity generating lattice vectors, $\ell_{_+}^{\,a}$ and  
$\ell_{_-}^{\,a}$ say, can be taken to be any two 
differences between pairs of such boundary segment vectors. For 
example, one can take
{\be \ell_{_\pm}^{\,a}= b_{_{(0)}}^{\,a}-b_{_\pm}^{\,a}
\, .\label{31}\fe}

\begin{figure}
\centering
\epsfig{figure=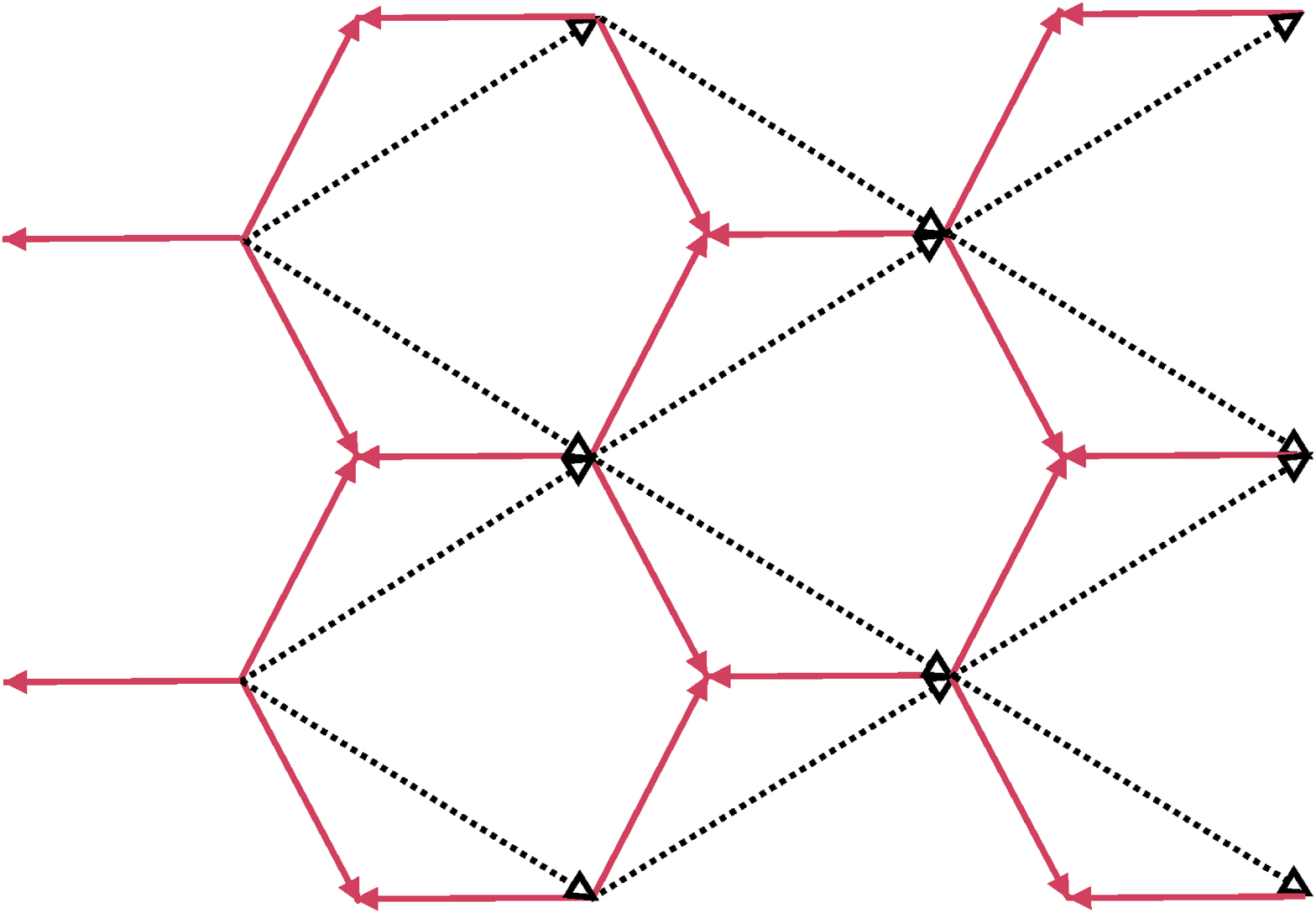, height=6 cm}
\caption{The dotted arrows indicate how the lattice periodicity
generators  $\ell_{_\pm}^{\,a}$ are placed in relation
to the material structure of a regular hexagonal lattice
of domain walls with even type Y-form junctions in 2
dimensions}
\label{Fig2}
\end{figure}

The fact that such a lattice is of odd type, with triply intersecting
Y-type junctions, means that when it is subject to a static 
macroscopically affine deformation, as specified by a linear 
transformation of the form
{\be \ell_{_\pm}^{\,a} \mapsto \overline  \ell_{_\pm}^{\,a} =
(\delta^a_b+\varepsilon^a_{\ b}) \ell_{_\pm}^{\,a} \, ,\label{32}\fe}
acting on the lattice periodicity generators, it is no longer
possible to suppose that this
affine transformation acts locally on the mesoscopic structure
characterised by the boundary segment vectors $b_{_{(0)}}^{\,a}$,  $b_{_+}^{\,a}$,  $b_{_-}^{\,a}$.
Thus, in order to find the new equilibrium state, which will allow us to compute the rigidity modulus, we need to find the transformations that, when applied to the boundary vectors ensure that all the periodicity vectors given by (\ref{31}), are subject to the same affine transformation (\ref{32}). Although this is a difficult problem in general, there is a simple solution in the particular case of the hexagonal lattice. This involves changing the lengths, but not the directions, of these boundary segment vectors, according to
\be b_{_{(0)}}^{\,a} \mapsto\overline b_{_{(0)}}^{\,a} =
(1+\varepsilon_{_{(0)}})  b_{_{(0)}}^{\,a}\, ,\hskip 1 cm
b_{_\pm}^{\,a}\mapsto \overline b_{_\pm}^{\,a}=
(1+\varepsilon_{_\pm})b_{_\pm}^{\,a}\, .\label{33}\fe
These transformations, which obviously preserve the equilibrium condition, will give rise to a fractional density change that will be given by 
\be\frac {\delta\rho}{\check\rho}=\frac{\varepsilon_{_{(0)}}+
 \varepsilon_{_+}+ \varepsilon_{_-}}{3}\, .\label{34}\fe

The effect of the non-affine transformation (\ref{33}) on the lattice 
vectors specified by (\ref{31}) can be expressed as an affine 
transformation of the form (\ref{32}), where the deformation matrix 
components are given by
\be \varepsilon^{_1}_{\ _1}=\frac{4\varepsilon_{_{(0)}}+
 \varepsilon_{_+}+ \varepsilon_{_-}}{6}\, ,\hskip 1 cm
 \varepsilon^{_2}_{\ _2}=\frac{\varepsilon_{_+}+ \varepsilon_{_-}}{2}
\, ,\hskip 1 cm \varepsilon^{_2}_{\ _1}=\varepsilon^{_1}_{\ _2}=
\frac{\varepsilon_{_+}-\varepsilon_{_-}}{2\sqrt3}\, .\label{35}\fe

In order to evaluate the components of the rigidity tensor we only need
to consider transformations that conserve volume, a condition which can be 
expressed by the requirement that the determinant of the affine 
transformation matrix should have unit value. This implies the restriction
\be \varepsilon_{_{(0)}}+\varepsilon_{_+}+ \varepsilon_{_-}=
-\frac{1}{2}\left(\varepsilon_{_{(0)}}\varepsilon_{_+}+
\varepsilon_{_{(0)}}\varepsilon_{_-}+\varepsilon_{_+}\varepsilon_{_-}
\right)\, .\label{36}\fe
This condition enables us to eliminate $\varepsilon_{_{(0)}}$ and work 
just with the two remaining independent parameters $\varepsilon_{_+}$
and $\varepsilon_{_-}$, thereby expressing the deformation matrix, to 
first order in the deformation amplitude, in the form
{\be \varepsilon^a_{\ b}=s^a_{\ b}+{\cal O}\{\varepsilon^2\}
\,.\label{37}\fe}
Here, $s^a_{\ b}$ is the trace free matrix with components given by
{\be  s^{_1}_{\ _1}=- s^{_2}_{\ _2}
=-\frac{\varepsilon_{_+}+ \varepsilon_{_-}}{2}\, ,\hskip 1 cm 
s^{_2}_{\ _1}=s^{_1}_{\ _2}=
\frac{\varepsilon_{_+}-\varepsilon_{_-}}{2\sqrt3}\, .\label{38}\fe}
In terms of these shear matrix components it can be seen that the 
fractional density variation (\ref{34}) can be written as
{\be\frac {\delta\rho}{\check\rho}=\frac{1}{4}\left( s^{_1}_{\ _1}{^2}+
s^{_2}_{\ _2}{^2}+2s^{_2}_{\ _1}s^{_1}_{\ _2}\right)\, ,\label{39}\fe}
which is manifestly of the isotropic form (\ref{05}) with a rigidity
coefficient that can be deduce to be 
{\be \check\mu=\frac{\check\rho}{4}\, .\label{40}\fe}

In the discussion above, we chose transformations that would preserve equilibrium and would be consistent with the periodicity vectors being subject to the same affine deformation, that is, working from the mesoscopic to the macroscopic scale --- a bottom-to-top approach. We should point out that we could also have done the calculation the other way around in a top-bottom approach. In this case, the problem is posed in the following way: consider a Y-junction and subject all of its points (excluding the vertex) to a given deformation and then look for the position of the vertex which, keeping the new endpoints of the boundary vectors fixed, minimizes the energy. If we consider the origin and the points $b^{a}_{(0)}$, $b^{a}_{\pm}$ as defining the Y-type junction then a transformation of the $b^{a}$s under $x\rightarrow (1+\varepsilon)x$ and $y\rightarrow y/(1+\varepsilon)$ and the origin moving to the point $(x,y)$ results in an energy functional
\begin{eqnarray}
E(x,y)&=&\sqrt{\left({1\over 2}(1+\varepsilon)-x\right)^2+\left({\sqrt{3}\over 2(1+\varepsilon)}-y\right)^2}+\sqrt{\left({1\over 2}(1+\varepsilon)-x\right)^2+\left({\sqrt{3}\over 2(1+\varepsilon)}+y\right)^2}\nonumber\\ &+& \sqrt{\left(1+\varepsilon-x\right)^2+y^2}\,.
\end{eqnarray}
One can then compute the position $(x,y)$ which minimizes $E(x,y)$ either numerically or, since we only need the positions to order $\delta^2$, as a power series. We find that $x=\varepsilon+\varepsilon^2/2+{\cal O}\{\varepsilon^3\}$ and $y={\cal O}\{\varepsilon^3\}$ are the minima and so $\Delta\rho/\check\rho=1/2$ and we conclude the that $\check\mu/\check\rho=1/4$. One can also consider a simple shear transformation $x\rightarrow x+2\varepsilon y$, $y\rightarrow y$ which yields
\begin{eqnarray}
E(x,y)&=&\sqrt{\left({1\over 2}+\sqrt{3}\varepsilon -x\right)^2+\left({\sqrt{3}\over 2}-y\right)^2}+\sqrt{\left({1\over 2}-\sqrt{3}\varepsilon -x\right)^2+\left({\sqrt{3}\over 2}+y\right)^2}\nonumber\\&+&\sqrt{(1+x)^2+y^2}\,,
\end{eqnarray}
which is minimized by $x=2\varepsilon^2+{\cal O}\{\varepsilon^3\}$ and $y=\varepsilon+{\cal O}\{\varepsilon^3\}$ and yields the same result. While the bottom-top approach is elegant, it is much more complicated when dealing with 3 dimensional lattices and is often intractable. As we shall see in section \ref{octdod}, in most cases the top-to-bottom approach is more practical, but even then it is often necessary to resort to numerical means to solve for the minimum point.

\begin{figure}
\centering
\epsfig{figure=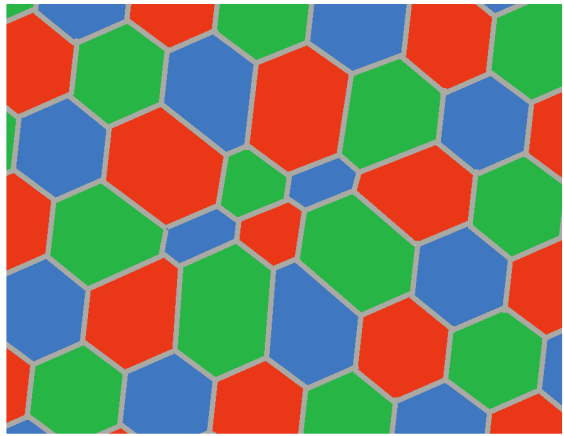, height=6 cm}
\caption{Locally confined energy conserving, and therefore linearly destabilising, deformation in ``odd'' type hexagonal lattice of domain walls  with Y-type junctions in 2 dimensions.}
\label{Fig3}\end{figure}

It can be seen from (\ref{25}) that $\check\beta=-\rho/4$ for a $D=2$ string/wall system and therefore using (\ref{23}) the value (\ref{40}) obtained for the rigidity modulus of this odd type hexagonal lattice has exactly the critical value for stability of longitudinal modes. According to  (\ref{21}) and (\ref{23}) the squared velocities of the transverse and longitudinal modes will be given respectively by
\be \upsilon_{\rm T}^{\, 2}=\frac{_1}{^2} \, ,\hskip 1 cm
\upsilon_{\rm L}^{\, 2}=0\, .\label{41}\fe
This contrasts with the corresponding case (\ref{29})
of the even type triangular lattice for which, the value (\ref{28}) of the 
rigidity modulus was found to be 50 percent larger, and so safely within 
the range required for absolute stability.

It is instructive to compare this hexagonal lattice case, for which the 
marginal zero velocity modes are of (irrotational) longitudinally 
polarised type, with the case of a simple perfect fluid for which, as 
remarked above, the marginal zero velocity modes are of transversely 
polarised (density conserving) type. The latter are not sufficient for 
local destabilisation of the fluid. However, it can be seen that the hexagonal 
lattice model really will be linearly, but not exponentially 
destabilised by its zero velocity longitudinal modes since they exist for all wavenumber directions, they can form a sufficient basis for superposition to constitute irrotational wave packets with locally confined support. The mesoscopic structure of an example of such a locally confined, zero energy, and therefore linearly destabilising, perturbation in a regular hexagonal lattice is shown in
Figure \ref{Fig3}. This shows the surprising result that although duality preserves the isotropic behaviour, it is not so for the stability.

\section{Orthonormal lattices}

\begin{figure}
\centering
\epsfig{figure=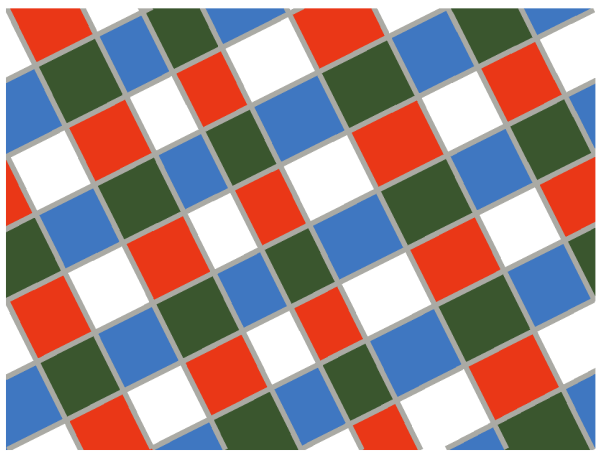, height=6 cm}
\caption{
Innocuous -- albeit energy conserving -- deformation, of axially aligned
plane wave  plane wave type, in square lattice with even type X form
junctions in 2 dimensions. Unlike the example in Figure \ref{Fig3},
such a perturbation is not locally confined, since it can be seen to be
incompatible with fixed boundary conditions at the upper and lower
borders of the figure, so it does not engender any local instability.}
\label{Fig4}\end{figure}

We now turn our attention to orthonormal lattices which in $D=2$ correspond to lattices with square symmetry and in $D=3$ to those with cubic symmetry. The analysis of such cases is facilitated by the fact that their
junctions are of even type -- more specifically their walls meet
at X type crossovers -- which means the effect of a static
 macroscopically affine deformation can be represented by an
affine transformation even at a local mesoscopic level. In particular this is because the X-type junctions are stable for any opening angle, as opposed to the very specific angle ($120^{\circ}$) in the hexagonal case.

To actually evaluate the relevant Cartesian rigidity components for the 
orthonormal lattices, we consider the 
effect of affine transformations of the general form 
{\be x^i\mapsto\overline x^i=x^i+\varepsilon^i_{\ j}x^j
\, ,\label{46}\fe} 
as specified by a deformation matrix $\varepsilon^i_{\ j}$ whose 
symmetrized trace free part,
{\be s^i_{\ j}=\frac{\varepsilon^i_{\ j}+\varepsilon^j_{\ i}}{2}
-\frac{\varepsilon^k_{\ k}\delta^i_j}{\srm D}\, ,\label{47}\fe}
can be interpreted as the Cartesian representation of the shear 
tensor given by (\ref{02}).

To evaluate the mixed component, $ {\mit\Sigma}^{_{1\,2\,1\,2}}$,
consider the volume conserving simple shear deformation for which 
the only non vanishing component of the deformation matrix is 
$\varepsilon^{_1}_{\ _2} =2\varepsilon$  say, so that the only non 
vanishing shear components of the corresponding shear matrix will be 
$s^{_1}_{\ _2}=s^{_2}_{\ _1}=\varepsilon$. This transformation has no 
effect on any of the relevant (originally) orthonormal basis vectors 
except the first, whose length will change 
by an amount $\delta\ell=2\varepsilon^2$, thereby bringing about a 
fractional change $\delta\rho/\check\rho=2\varepsilon^2/{\srm D}$ in 
the mean density. Since the effect of the corresponding shear 
deformation (\ref{47}), whose only non-vanishing matrix components 
are $s^{_1}_{\ _2}=s^{_2}_{\ _1}=\varepsilon$ will be given according
 to (\ref{01}) by $\delta\rho=2{\mit\Sigma}^{_{1\,2\,1\,2}}\,
\varepsilon^2$, the required shear component value can be deduced to be 
{\be {\mit\Sigma}^{_{1\,2\,1\,2}}=\frac{\check\rho}{\srm D}
\, .\label{50}\fe}

To evaluate the other independent component, consider the irrotational 
volume conserving shear deformation for which the only non vanishing 
components of the deformation matrix are given to quadratic order by
$\varepsilon^{_1}_{\ 1}=\varepsilon$, $\varepsilon^{_2}_{\ 2}
=-\varepsilon+\varepsilon^2$. In this case there are two (originally) 
orthonormal basis vectors whose length is affected, namely the first, 
for which $\delta\ell=\varepsilon$, and the second, for which 
$\delta\ell=-\varepsilon+\varepsilon^2$. The combined effect of these 
length adjustments evidently cancels out to first order, producing a 
fractional density change that is given at quadratic order by 
$\delta\rho/\check\rho=\varepsilon^2/{\srm D}$. In this case the only 
non-vanishing shear matrix components will be given to first order by 
$s^{_1}_{\,1}=-s^{_2}_{\,2}=\varepsilon$, so the corresponding
density change will be given according to (\ref{02}) by
$\delta\rho/\check\rho=({\mit\Sigma}^{_{1\,1\,1\,1}}-
{\mit\Sigma}^{_{1\,1\,2\,2}})\,\varepsilon^2$. It follows 
from (\ref{45}) that the required values can be read out as
\be {\mit\Sigma}^{_{1\,1\,1\,1}}=\frac{({\srm D}-1)\check\rho}{\srm D^2}
\hskip 1 cm  {\mit\Sigma}^{_{1\,1\,2\,2}}=
-\frac{\check\rho}{\srm D^2}\, .\label{51}\fe

The outcome, therefore, is that for a lattice of the orthonormal
kind, the  rigidity coefficients introduced 
in (\ref{15}) will be given by
\be \check\mu_{\rm T}=\frac{\check\rho}{\srm D}\ ,\hskip 1 cm
\check\mu_{\rm L}=\frac{\check\rho}{2\srm D} \, .\label{13}\fe
Using our earlier analysis, it is clear that the longitudinal rigidity is only sufficient for those modes to be a marginally stable. We can also deduce, the scalar shear modulus
\be 
{\check\mu\over\check\rho}=\sqrt{2D+1\over 2D^2(D+2)}\,,
\fe
and the anisotropy parameter is 
\be 
{\Delta\check\mu\over\check\rho}={1\over 2D}\,.
\fe
For $D=2$, we have that $\check\mu/\check\rho\approx 0.395$ and $\Delta\check\mu/\check\rho\approx 0.250$, whereas for $D=3$ one can deduce that $\check\mu/\check\rho\approx 0.279$ and $\Delta\check\mu/\check\rho\approx 0.167$.

When used in conjunction with the formula (\ref{25}) for
the bulk modulus, it can be seen that the relations (\ref{13})
lead to a dramatic simplification of the matrix $Q^{ab}$
in the characteristic equations (\ref{18}) as given by the
of the formulae (\ref{qii}) and (\ref{qij}) in which most of
the terms cancel out, leaving
\be Q^{_{ii}}=\frac{\check\rho}{\srm D}\,{\sin}^2\theta_{_i}\, ,
\hskip 1 cm  Q^{_{ij}}=0\, .\label{63}\fe
This implies that the matrix $Q^{ab}$ will be of purely diagonal
form, which implies that, unlike the isotropic case, 
the allowed directions of the polarisation eigenvector 
$\iota^a$ in (\ref{18}) are independent of the propagation direction,
being fixed to coincide with the directions of the preferred 
orthonormal axes. On the other hand, still unlike what occurs in 
isotropic cases, the eigenvalues and hence the propagation speed 
will depend on its direction. 

In particular, by substituting (\ref{63}) into (\ref{18}), it can 
be seen, for example, that the mode with $\iota^a$ aligned with the
$x^{_i}$ axis will propagate with a value $\upsilon_{_{(i)}}$
say that will be given by
\be\upsilon_{_{(i)}}^{\,2}=\frac { {\sin}^2\theta_{_i}}{\gamma
{\srm D}}\, ,\label{64}\fe
for $i=1,..,D$ where $\gamma$ is the relevant polytropic index, which
according to (\ref{26}) will be given by $\gamma=1/{\srm D}$ for
a lattice of cosmic domain walls and by $\gamma=({\srm D}-1)
/{\srm D}$ for the other case to which the present analysis applies, 
namely a lattice of cosmic strings.  

In the physically interesting case of domain walls, independent of the 
dimension we end up with the memorable formula
\be\upsilon_{_{(i)}}^{\,2}={\sin}^2\theta_i
\, ,\label{65}\fe
which tells us that the velocity ranges up to the speed of light in 
the limit of propagation orthogonal to the polarisation direction, and 
down to zero for propagation parallel to the polarisation direction.  
In the more artificial case of an orthonormal string lattice the 
velocity will also range down to zero for propagation parallel to the 
polarisation direction, but (except in the 2 dimensional case 
where the distinction between walls and strings does not arise) the 
maximum velocity possible for propagation orthogonal to the 
polarisation direction will be reduced by a factor $1/\sqrt{{\srm D} -1}$. 

Unlike the highly stable example of the  triangular lattice,
which according to (\ref{29}) has no zero velocity modes at all,
the orthonormal lattices considered here
do have some exceptional zero velocity  
unstable plane wave modes, namely those with an exactly vanishing
value of the angle in (\ref{64}) measuring the deviation of the 
propagation from the relevant preferred axis. However, as all the
neighbouring modes are strictly stable, the occurrence of these
exceptional modes is not sufficient for the construction of
unstable locally confined perturbations.  The situation in these
square and cubic examples is therefore similar to that in an
ordinary fluid, which has unconfined linearly unstable shear
modes, but is nevertheless effectively stable with respect to 
perturbations having confined support. These modes correspond to the kind of global deformation illustrated in Fig.~\ref{Fig4}

This contrasts with the effectively unstable case of the
hexagonal lattice  for which mesoscopically non-affine deviations reduce the 
energy associated with a macroscopically affine deviation to such an extent 
that, according to (\ref{41}) it possesses linearly unstable plane wave 
modes of longitudinally polarised type in all directions. Although such 
waves are of individually unconfined support, they are sufficiently 
numerous for their collective superpositions to provide 
 linearly unstable perturbations that are initially confined 
within a finite region, thereby effectively destroying the stability of
 the system as a whole.

\section{Primitive lattices in 3D with cubic symmetry}
\label{octdod}

We have already commented that the anisotropic forms of the elasticity tensor can be classified in 3D by the Bravais lattices. Since our ultimate aim is the application to the idea of Solid Dark Energy, it seems sensible to consider the three primitive lattices allowed in 3D which are the Wigner-Seitz cells of simple cubic (SC), body-centred cubic (BCC) and face-centred cubic (FCC) lattices. The polyhedra corresponding to these lattices are the simple cube, already studied in the previous section, the truncated octahedron (sometimes known as the the tetrakaidecahedron~\cite{thomson}) and the rhombic dodecahedron. The latter two fall into the class of Archimedean solids and are space-filling, as is the former which is a platonic solid. The truncated octahedron has 14 faces, 6 of which are squares and 8 are hexagons. The rhombic dodecahedra is a  polyhedron
with 12 faces each of which is a rhombus.

Unfortunately, the analytic techniques applied in the previous sections are intractable in these cases. However, since similar ideas have been studied extensively in the context of soap films, foams and other soft condensed matter~\cite{weaire}, a software package, known as the Surface Evolver~\cite{surface}, has been developed to minimize the surface area of polyhedra of arbitrary topology. This package, which is publicly available, allows one to accurately compute the energy of minimal energy configurations in an unperturbed state and ones which are subject to a small deformation.

To obtain the rigidity coefficients we select deformations
in an analogous fashion to the previous sections. 
The mixed component of the
rigidity tensor $ {\mit\Sigma}^{_{1\,2\,1\,2}}$ is evaluated by
considering the volume conserving simple shear deformation for which
the only non vanishing component of the deformation matrix is
$\varepsilon^{_1}_{\ _2} =2\varepsilon$; we will denote this DEF1 in the subsequent discussion. This will allow us to compute $\mu_{\rm T}$. To evaluate the other independent component, $\mu_{\rm L}$, 
we  consider the irrotational  volume
conserving shear deformation for which the only non vanishing
components of the deformation matrix are given to quadratic order by
$\varepsilon^{_1}_{\ 1}=\varepsilon$, $\varepsilon^{_2}_{\ 2}
=-\varepsilon+\varepsilon^2$; this will be denoted DEF2. In both cases we compute the fractional energy change $\delta\rho/\check\rho$ for a range of small values of $\varepsilon$ and fit for the coefficient of  $\varepsilon^2$.

\begin{figure}
\centering
\mbox{\resizebox{0.4\textwidth}{!}{\includegraphics{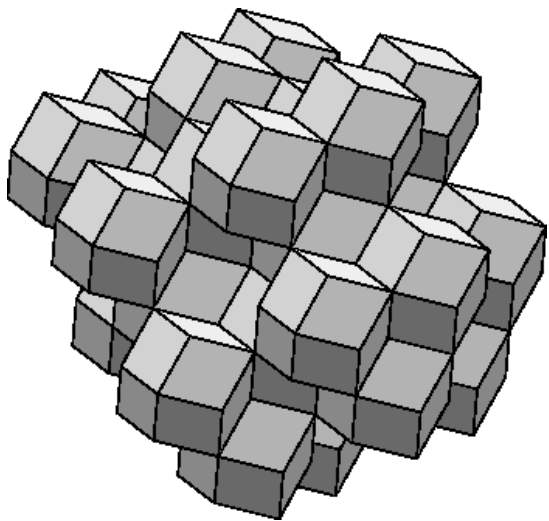}}}
\caption{A lattice of rhombic dodecahedra in an unperturbed minimal energy configuration. This configuration with flat faces is the minimal energy configuration.}
\label{rd_unpert}
\end{figure}

Let us first consider a lattice of rhombic dodecahedra in an FCC
 arrangement as illustrated in  Fig~\ref{rd_unpert}. The
 configuration has flat faces and by using Surface Evolver we find that
 this is the local minimum energy configuration with this topology.
This structure contains a
 significant fraction of odd-type junctions as well as some even type
 junctions which will help to stabilise the system as they will not be
 required to adjust themselves after an affine transformation - in a
 system with purely even-type junctions the non-affine correction is
 zero.

Allowing for only affine deformations with of the kind DEF1, we find that 
 the fractional change in the mean density is $\delta\rho/\check\rho= 0.5 \varepsilon^{2}$. Subsequent energy minimisation using the Surface Evolver gives
 $\delta\rho/\check\rho= 0.333 \varepsilon^{2}$. Using DEF2, we find that 
 $\delta\rho/\check\rho= 0.583 \varepsilon^{2}$ and
 $\delta\rho/\check\rho= 0.445 \varepsilon^{2}$ for the affine and
 non-affine cases, respectively. Using these values we find that 
\be {\check\mu_{\rm L}\over\check\rho}\approx0.222\,,\qquad {\check\mu_{\rm T}\over\check\rho}\approx 0.167\,,\fe
for this configuration when we take into account non-affine perturbations.
 Using higher numerical accuracy than the 3 decimal places quoted here, we deduce that it is likely that $\check\mu_{\rm L}/\check\rho=2/9$ and $\check\mu_{\rm T}/\check\rho=1/6$ although this can only be checked to a particular level of accuracy.

\begin{figure}
\centering
\mbox{\resizebox{0.3\textwidth}{!}{\includegraphics{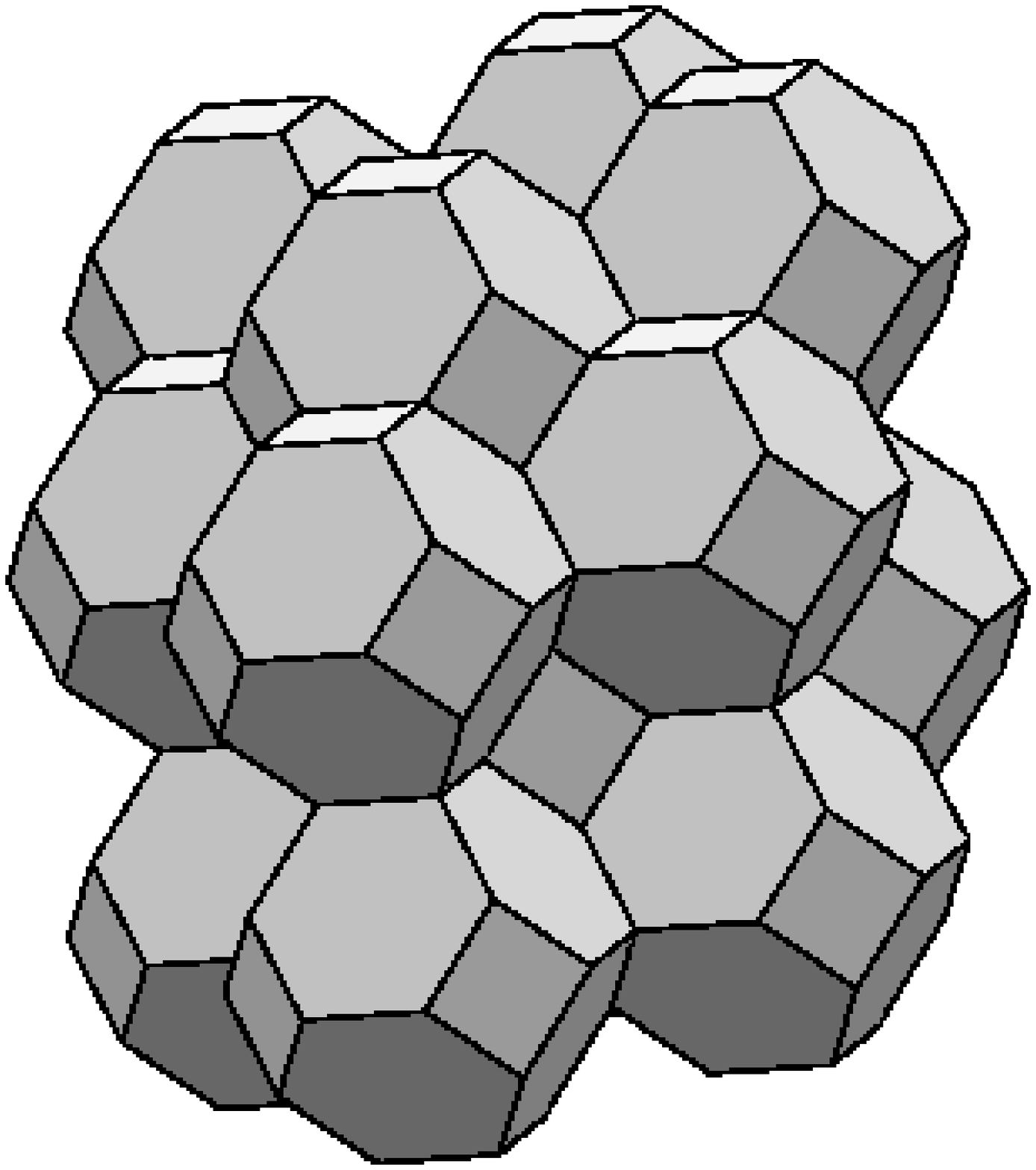}}}
\qquad\qquad\mbox{\resizebox{0.3\textwidth}{!}{\includegraphics{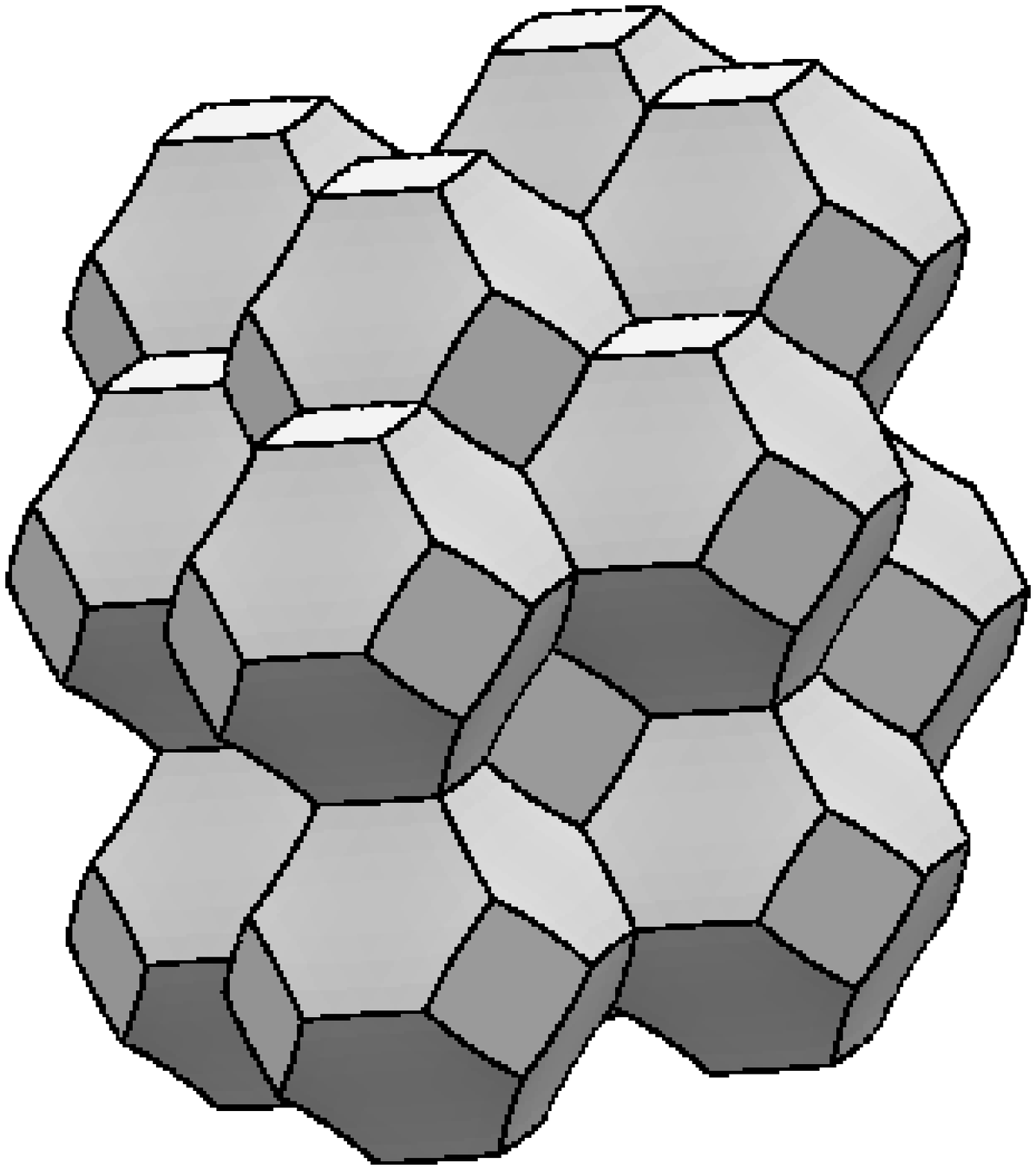}}}
\caption{On the left is an illustration of the BCC arrangement of truncated octahedra in unperturbed state and on the right is the unperturbed minimal energy configuration created using the Surface Evolver. Notice the slight curvature of the faces on the right.}
\label{bcc}
\end{figure}

Next we consider a lattice of truncated octahedra in BCC arrangement. This is the configuration which was proposed by Lord Kelvin as the polyhedron with the lowest surface area for a fixed volume --- the so called Kelvin Problem~\cite{thomson,weaire}. Although this is now known not to be the absolute minimum~\cite{weairephelan}, it is has been extensively studied in the context of soap films. As illustrated in Fig.~\ref{bcc}, truncated octahedra with flat faces are not the local minimum energy configuration: the minimum has curved faces.

Under an affine transformation DEF1 there is a fractional change in
the mean density of $\delta\rho/\check\rho= 0.494 \varepsilon^{2}$ for truncated octahedra with flat faces. A subsequent energy minimisation using the Surface Evolver allowing for non-affine transformations, reduces  the fractional change in the mean density which is given by $\delta\rho/\check\rho= 0.367 \varepsilon^{2}$. For DEF2, the fractional change in the  mean density is $\delta\rho/\check\rho= 0.592 \varepsilon^{2}$ for the affine deformation - subsequent vertex readjustment due to energy minimisation reduces this to
$\delta\rho/\check\rho= 0.218 \varepsilon^{2}$. Therefore, we find that for
the Kelvin foam of truncated octahedra the rigidity coefficients are
given by
\be {\check\mu_{\rm T}\over\check\rho}\approx 0.183\,,\qquad {\check\mu_{\rm L}\over\check\rho}\approx 0.109.
\fe  
These values are compatible with previous calculations done in the context of soap films~\cite{KR}.

\begin{table}
\begin{center}
\begin{tabular}{|l|c|c|c|c|}
\hline
   & $\check\mu_{\rm L}/\check\rho$ & $\check\mu_{\rm T}/\check\rho$ & $\check\mu/\check\rho$ & $\Delta\mu/\check\rho$ \\
\hline
SC  & 1/6 & 1/3 & $\sqrt{7/90}$ & 1/6 \\
FCC & 2/9 & 1/6 & $\sqrt{59/1620}$ & -1/18 \\
BCC & 0.109 & 0.183 & 0.150 & 0.074 \\
\hline
\end{tabular}
\end{center}
\caption{The shear moduli computed using for the simple cubic (SC), face-centred cubic lattices and body-centred cubic (BCC) lattices.}
\label{tab:tab1}
\end{table}

We have tabulated the compute shear moduli for the SC, FCC and BCC cells in table~\ref{tab:tab1} along with the computed scalar shear modulus $\check\mu/\check\rho$ and the anisotropy $\Delta\mu/\check\rho$. We see that the value of $\check\mu/\check\rho$ is lower than the value computed in ref.~\cite{BCCM05} due to the effects of non-affine deformation and also the strong anisotropy present in all cases. Using the results derived in section~\ref{anisostable} one can see that the SC and FCC cell structures are conditionally stable with zero modes in each case. These are longitudinal in the SC cases and transverse in the the FCC case. Moreover we find that the BCC lattice is unstable since $\mu_{\rm L}=0.109<1/6$.

\begin{figure}
\centering
\epsfig{figure=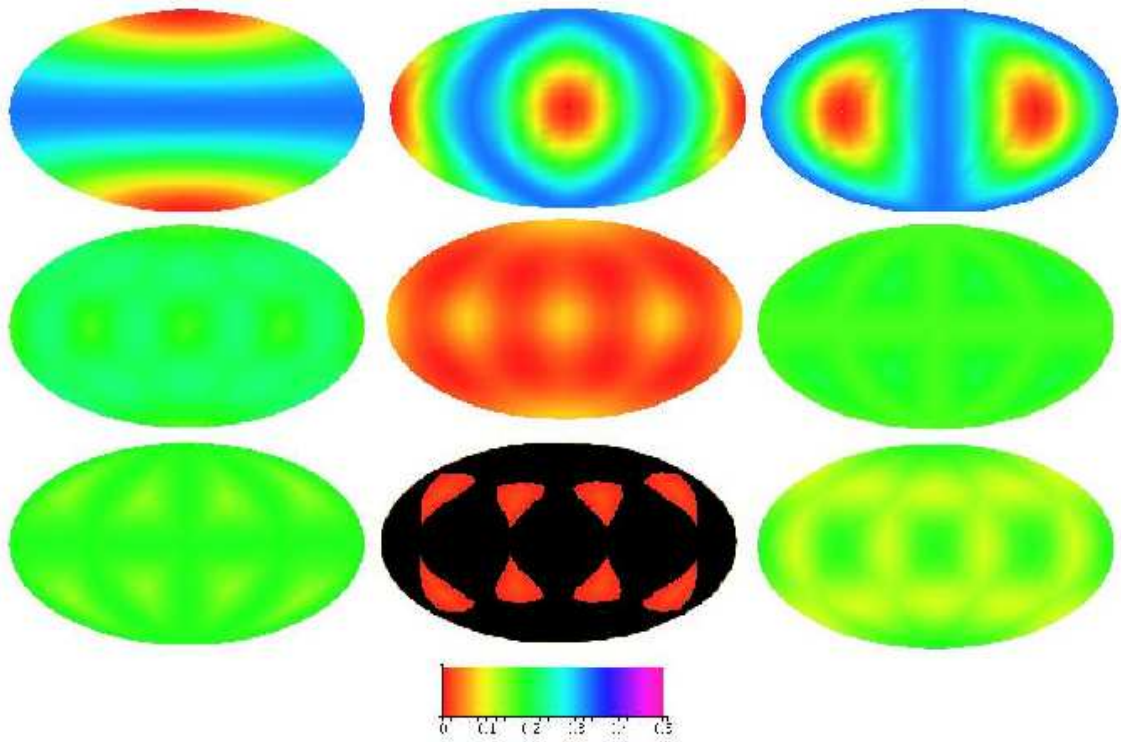,height=10cm,width=15cm}
\caption{The computed square sound speeds for the three orthogonal modes plotted as a function of the polar angles $(\theta,\phi)$ of Fourier space in the equal-area Hammer Aitoff project. The top three are for the simple cube, the middle three are for the lattice of rhombic dodecahedra and the bottom three are for the lattice of rhombic dodecahedra. Black on the colour bar denoted points where the square sound speeds are negative corresponding to exponentially unstable regions}
\label{allspeeds}
\end{figure}

It is not possible to compute the sound speeds analytically for the FCC and BCC cases as we did for the SC case in the previous section. In Fig.~\ref{allspeeds}, we have plotted the numerically computed square sound speeds $(\upsilon^2)$ as a function of $(\theta,\phi)$ for all three lattices in an equal area Hammer-Aitoff projection where ${\bf k}=k(\sin\theta\cos\phi,\sin\theta\sin\phi,\cos\theta)$. In all three cases the characteristic four-fold symmetry is evident. However, there are some interesting differences: for the BCC lattice there is a substantial region of the space $(\theta,\phi)$ for which $\upsilon^2<0$, confirming the instability already identified; the SC lattice has $\upsilon^2=0$ at the centre of each of the faces of a cube, whereas the FCC lattice has $\upsilon^2=0$ in directions exemplified by (\ref{nonaxial}).

\section{Discussion and conclusions}

In this paper we have examined the stability of a variety of different lattices both in 2 and 3 dimensions (some of our results even apply in arbitrary dimension). There are two features which we have investigated which go beyond our earlier analysis~\cite{BCCM05}. Firstly, we have confirmed the physical importance of the distinction between odd type and even type junctions between the relevant wall (or string) segments. It appears that the more odd type junctions a given lattice cell has the less stable it is likely to be. The other feature which we have highlighted are the effects of anisotropy. Even though the orthonormal lattices contain only even type junctions, the strong anisotropy present in these models is important.

Since our main aim was to find a lattice which is stable as a candidate for solid dark energy it is worth discussing progress toward this. The 3 dimensional configurations which we have investigated are formed from cells based around the simple cube, rhombic dodecahedra and the truncated icosahedron. These are the primitive Bravais lattices with cubic symmetry.

Our calculations suggest that we should immediately eliminate the cell based on the truncated octahedron which is exponentially unstable for a wide range of $(\theta,\phi)$. This is something of a surprise since it is known at least play a role in the structure of soap films. However, this highlights an important difference between that field and the one at issue here. Namely the bubbles in soap films have an extra contribution to the bulk modulus due to the air pressure in the bubble, whereas we require that the domain wall lattices exists in vacuum.

We have shown that the rhombic dodecahedral lattice exhibits transverse zero modes. The question of whether these zero modes are of infinite extent or have local support is more difficult as the junctions are of mixed type. As there are eight principle directions for which there is a zero mode we conjecture that it is possible to construct a closed path around these directions (as in the hexagonal case). In the hexagonal case the length scale of this finite mode is comparable to the elementary cell size. However, due to the existence of the even-type junctions in the rhombic dodecahedron the length scale of the mode may potentially be large in comparison to the elementary cell. We therefore conclude that this mode will be of larger extent than in the hexagonal case, although this should be tested numerically. Indeed, as described below, this will be the subject of some future work.

Remarkably, the best candidate that we have at present is the simple cubic lattice which has only has unstable modes which are of infinite extent. These zero modes correspond to there being no specific distance fixed between infinite straight walls with the model based on walls with Nambu-Goto equation of state. Clearly this issue, along with the precise spatial distribution of the vacua, would need to be addressed by any field theoretical model. We are presently investigating this issue in an $O(N)$ model with cubic anisotropy and the results will be presented in a future publication~\cite{BCM}.

One should not take our comments above to reflect pessimism about the Solid Dark Energy scenario; this analysis is only the beginning of a search for stable configuration. There are number of possibilities for further investigation: in particular, the other primitive Bravais lattices with lower symmetry than cubic and compound structures with cubic symmetry such as the tetrahedral close packing structures. We plan to investigate these possibilities in future work. 

\section*{Acknowledgements}
\noindent The authors wish to thank Martin Bucher and Brandon Carter for stimulating discussions.

\end{document}